# Data-Efficient Automatic Shaping of Liquid Droplets on an Air-Ferrofluid Interface with Bayesian Optimization

P. A. Diluka Harischandra and Quan Zhou

*Abstract—* Manipulating the shape of a liquid droplet is essential for a wide range of applications in medicine and industry. However, existing methods are typically limited to generating simple shapes, such as ellipses, or rely on predefined templates. Although recent approaches have demonstrated more complex geometries, they remain constrained by limited adaptability and lack of real-time control. Here, we introduce a data-efficient method that enables real-time, programmable shaping of nonmagnetic liquid droplets into diverse target forms at the air-ferrofluid interface using Bayesian optimization. The droplet can adopt either convex or concave shapes depending on the actuation of the surrounding electromagnets. Bayesian optimization determines the optimal magnetic flux density for shaping the liquid droplet into a desired target shape. Our method enables automatic shaping into various triangular and rectangular shapes with a maximum shape error of 0.81 mm, as well as into letter-like patterns. To the best of our knowledge, this is the first demonstration of real-time, automatic shaping of nonmagnetic liquid droplets into desired target shapes using magnetic fields or other external energy fields.

## I. INTRODUCTION

Shaping liquid droplets into controlled, non-equilibrium configurations is a fundamental challenge in fluid dynamics, with applications in printing, coating, optics, and bioengineering [1], [2], [3]. Various techniques have been explored for droplet manipulation, including template-based approaches, energy-based shaping methods, and ferrofluid-based systems. However, most existing methods remain constrained by material dependencies, shape limitations, or the need for complex fabrication, making real-time, programmable droplet shaping a largely unsolved problem.

Template-based methods offer a straightforward approach, where droplets conform to a predefined mold [12]. While these methods enable controlled shaping, they lack adaptability, as the droplet's final shape is entirely dictated by the template. As a result, they are unsuitable for applications requiring dynamic shape reconfiguration or real-time control.

Several external energy fields such as airflow [4], acoustic waves [5], electric fields [6], [7], and thermal gradients [8], [9], [10] have also been employed to manipulate liquid droplets. These active methods provide more flexibility, but typically require the droplet to exhibit specific material properties such as conductivity, dielectric response, or sensitivity to acoustic radiation forces, limiting their applicability to a narrow class of liquids. Electrically induced shaping, for example, can generate complex droplet morphologies [11], but it depends on precise tuning of the liquid's electrical properties and external voltages, limiting its general applicability.

Ferrofluid-based droplet manipulation introduces a distinct category of techniques that leverage a magnetically responsive liquid as the actuation medium. Some approaches use a sheet of ferrofluid wrapped in polytetrafluoroethylene (PTFE) films to indirectly move and mix liquid droplets using magnetic fields [13], but this method suffers from fabrication complexity and lacks shape control. Other methods use magnetic fields to directly deform ferrofluid droplets into elongated, elliptical [14], [15], or ring-shaped configurations [16]. Recently, magnetic shaping of liquid metal infused with iron nanoparticles has also produced cross-shaped patterns [17]. Despite these advances, such methods inherently require the droplet to be magnetically responsive, restricting their broader application.

To overcome these limitations, the authors previously demonstrated electromagnetic shaping of nonmagnetic liquid droplets at the air–ferrofluid interface [18]. This method achieved remarkable shape manipulation, producing elliptical, concave, and convex forms, as well as enabling droplet rotation and shaping of phase-change droplets. However, despite its versatility, the approach relied on predefined open-loop actuation, requiring extensive manual tuning for each shape and lacking real-time adaptability. Building on this, the authors later explored learning-based control using long short-term memory (LSTM) networks to predict droplet shape evolution [19]. While this introduced predictive capabilities, it required large-scale training data and was fundamentally open-loop in nature, lacking adaptive real-time optimization and limiting its practicality.

In this paper, we propose a data-efficient, real-time method for shaping nonmagnetic liquid droplets at the air-ferrofluid interface. Our approach leverages Gaussian processes and Bayesian optimization to estimate optimal solenoid actuation, dynamically deforming the interface to achieve physically realizable target shapes without requiring predefined templates

*Research supported by Academy of Finland projects 331149 and 349532, and Finnish Automation Foundation.

P. A. Diluka Harischandra and Quan Zhou are affiliated with the Department of Electrical Engineering and Automation, School of Electrical Engineering, Aalto University, 02150 Espoo, Finland (e-mail: diluka.harischandra@gmail.com, quan.zhou@aalto.fi)

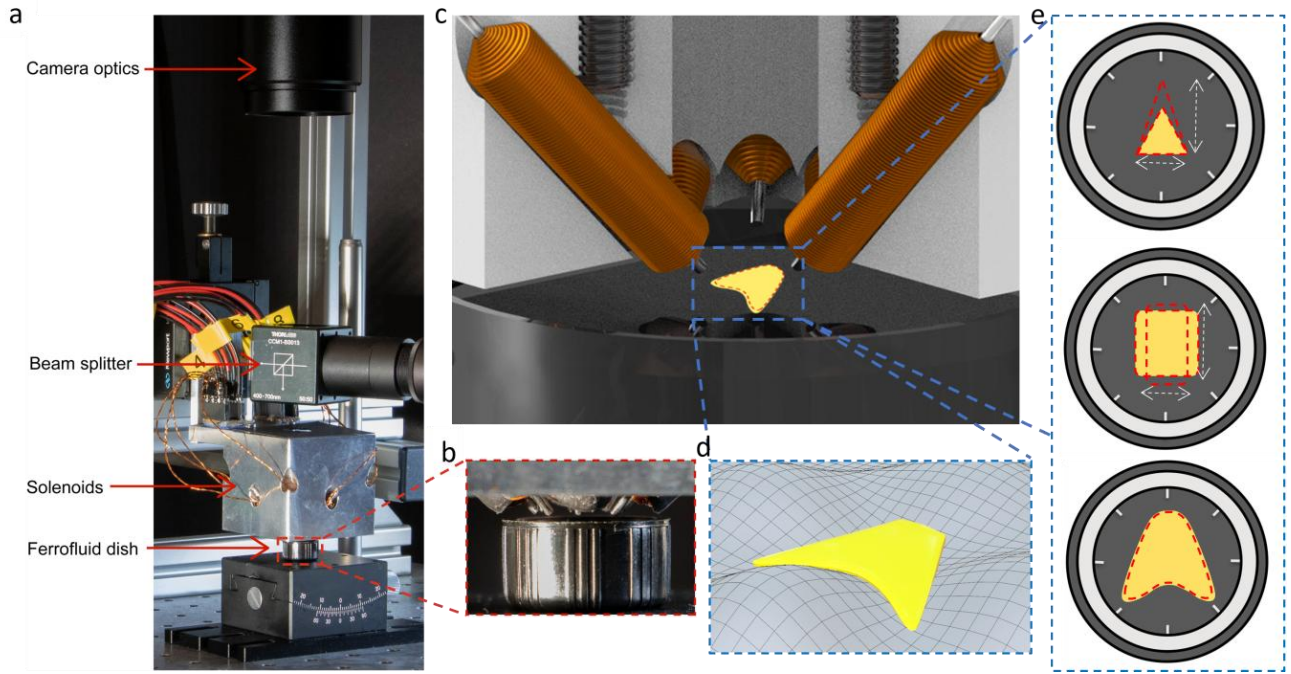

Figure 1. Concept of shaping liquid droplets on an air-ferrofluid interface. a) Experimental setup. b) Close-up view of the ferrofluid dish and the solenoid tips. c) Illustration of the solenoids and the shaped liquid at the air-ferrofluid interface. d) Illustration of the deformed air-ferrofluid interface and the shaped liquid. e) Example shapes of the droplet shaping results.

or extensive datasets. Unlike previous methods, this approach enables real-time shape adaptation within a few actuation steps while extending beyond simple elliptical deformations.

This work demonstrates, for the first time, real-time and programmable shaping of nonmagnetic liquid droplets using magnetic actuation, without relying on predefined templates or material-specific droplet properties. The rest of the paper is organized as follows. Section II describes the experiment setup, shape encoding, software implementation, and the optimization methods. Section III reports automatic shaping results. Finally, section IV concludes the article.

## II. METHODS

### A. Experimental setup

The concept of shaping liquid droplets on an air-ferrofluid interface is shown in Figure 1. The experimental setup, shown in Figure 1a, is adapted from our previous work on droplet manipulation and shaping liquid at the air-ferrofluid interface [18]. A close-up view of the ferrofluid dish is shown in Figure 1b. The dish has a volume of 1.36 mL, a diameter of 1.475 cm, and a depth of 0.795 cm.

Eight electromagnetic solenoids are used to deform the air-ferrofluid interface. Each of the solenoids comprises 300-400 turns of SWG 18 copper wire wound around a 1 mm diameter martensitic steel core. The solenoids are arranged at a 45° inclination from the horizontal plane and are spaced 45° apart, as illustrated in the CAD model in Figure 1c. Solenoids are calibrated using a Hall sensor (SS495A1, Honeywell, USA) and actuated with eight current controllers (ESCON 50/5, Maxon Group, Switzerland).

To minimize the vibrations at the air-ferrofluid interface, ramped analog waveforms are generated using an analog output device and are provided as reference signals to the current controllers. The workspace is illuminated using a collimated LED source (MNWHL4, Thorlabs Inc.) through a 50:50 beam splitter (CCM1-BS013, Thorlabs Inc.).

Commercial ferrofluid (EMG 408, Ferrotec, USA) is used, diluted with MQ water at 1:7 ratio. The shaped liquid is a 0.2 µL hexadecane droplet dispensed at the air-ferrofluid interface.

### B. Working principle

The actuation of solenoids creates bumps at the air-ferrofluid interface, as shown in the 3D illustration in Figure 1d. These bumps deform the hexadecane droplet. The working mechanism can be described using the total energy [18]:

$$E_T = E_{AF} + E_{OA} + E_{OF} + E_G + E_M \quad (1)$$

where, $E_{AF}, E_{OA}, E_{OF}$ denote the interfacial energies of the air-ferrofluid interface, oil-air interface, and the oil-ferrofluid interface. $E_G$ denotes the gravitation potential energy and $E_M$ denotes the magnetic energy. The gravitation potential energy depends on the effective mass of the droplet, defined as $m_{eff} = \varrho_O V_O - \varrho_F V_{imm}$ and the deformation of the ferrofluid interface induced by the magnetic field. The density and volume of the droplet are represented by $\varrho_O$ and $V_O$, respectively. $\varrho_F$ and $V_{imm}$ represent the density of the ferrofluid and the volume of the ferrofluid that the oil droplet displaces, respectively.

Oil droplets behave similarly to high-density particles at the air-ferrofluid interface [20], tending to move away from the bump created by interface deformation. This is attributed to the positive effective mass [18], resulting from the similar densities of the hexadecane droplet (0.77 gcm$^{-3}$) and the ferrofluid (0.99 gcm$^{-3}$).

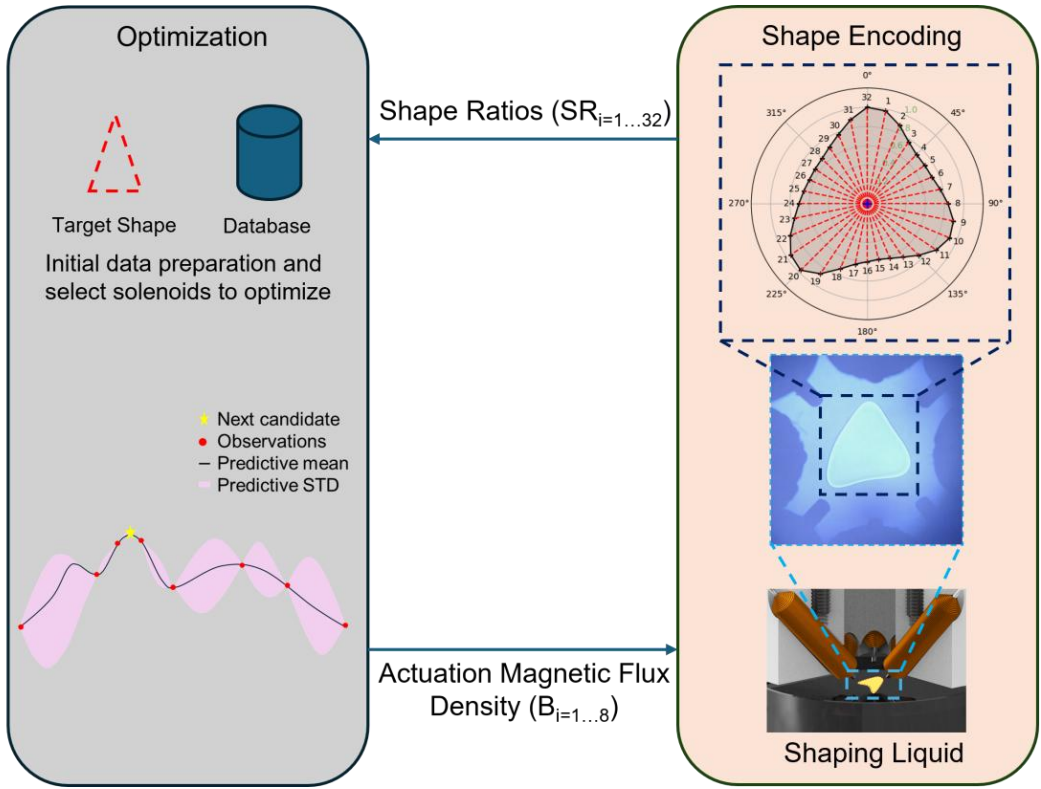

Figure 2. Automatically shaping a droplet into a target shape with data-efficient Bayesian optimization method. Initially, a random dataset of responses is prepared, with data corresponding to the selected actuation solenoids based on empirical data. A Bayesian optimizer generates a batch of five experiment candidates. The multicoil controller executes these experiments, tracking the droplet with a blob tracker. The droplet's contour is encoded into 32 angle-based segments. The shape encoded information from the experiments is then fed back to the Bayesian optimizer to generate new actuation candidates.

By actuating the solenoids with different magnetic flux densities at the tip, various convex and concave shapes of the hexadecane droplet can be achieved. Figure 1e illustrates the concept of shaping droplets into triangles and rectangle shapes of different aspect ratios, as well as letter shapes. A hexadecane droplet at the air-ferrofluid interface has a positive spreading coefficient, which cause it to spread at the interface [18]. Table I summarizes the properties of the utilized liquids.

TABLE I. PROPERTIES OF THE MANIPULATION LIQUIDS

| Liquid | Surface Tension (mN m$^{-1}$) | Density (gcm$^{-3}$) | Viscosity (mPa·s) |
|---|---|---|---|
| Ferrofluid | 70.88 | 0.99123 | 3.558 |
| Hexadecane | 27.39 | 0.76851 | 1.138 |

*C. Problem formulation*

In this study hexadecane droplets are shaped using a programmable air-ferrofluid interface, where the actuation of the solenoids determines the shape of the droplet, ranging from convex to concave. To enable automatic shaping, the droplet shape must first be represented in a form that can describe both convex and concave shapes. Secondly, it is desirable to reach the target shape as quickly as possible, since the hexadecane droplet spreads at the interface over time.

Unlike many physical systems that can be readily simulated, this case involves a complex three-phase system (air, oil, ferrofluid), which would require full 3D Finite Element Method (FEM) simulations, an approach that is computationally intensive and time-consuming. Therefore, we rely on empirical methods. However, once an oil droplet is placed at the air-ferrofluid interface, it cannot be removed, and a new ferrofluid dish must be used after the droplet has fully spread. This makes the experiments both costly and time-consuming.

Thus, a method that converges with few iterations is highly desirable. Bayesian optimization is particularly well suited for this purpose due to its efficiency in finding optimal solutions with minimal evaluations. Figure 2 illustrates the overall implementation of the automatic shaping method. This approach is purely data-driven and does not depend on a theoretical model. The implementation is described in detail in the following sections.

*D. Shape encoding*

The shape of a hexadecane droplet at the air-ferrofluid interface can be either convex or concave, depending on the programmable actuation of the solenoids [18]. The droplet is tracked using ViSP blob tracker [21] and its contour is extracted from the tracked blob.

The shape is encoded using 32 angle-based segments ($SR_i, i = 1 \ldots 32$) along the droplet boundary, following the approach from our previous work on learning the shape evolution under sequential actuation [19]. The distance from the center of the droplet to each segment is measured and normalized by dividing by the maximum distance, resulting in a size-invariant measure of the shape, referred to as shape ratio $SR_i$.

To achieve optimal results, it is essential to maintain a balance between precision and computational efficiency, especially given the time-varying nature of the air-ferrofluid interface, the behavior of the hexadecane droplet, and the processing time required by Bayesian Optimization. While increasing the number of segments along the droplet boundary improves ability to replicate complex target shapes, it also increases the complexity of optimization. Conversely, using fewer segments accelerates the optimization process, but it can introduce aliasing errors and hinder accurate representation of the desired shape.

*E. Software implementation*

The machine vision and the solenoid actuation processes are handled by in-house developed C++ software. A TCP server module is implemented using Boost TCP library. The Bayesian optimizer is implemented on the client side using BoTorch [22] and PyTorch libraries.

The Bayesian optimizer on the client side requests new experiment tests from the server by sending JSON-encoded magnetic flux density ($B_i, i = 1 \ldots 8$) for the actuation of the eight solenoids. The server receives the actuation values, performs the experiment, and returns the resulting JSON-encoded shape ratios $SR_i, i = 1 \ldots 32$.

*F. Shape optimization*

A Gaussian Process (GP) was employed to model the shape response to solenoid actuation, using the BoTorch libraries [22]. The GP takes eight inputs representing the actuation currents, and 34 outputs representing 32 shape ratios and the center coordinates of the shape. The model is trained using a Matérn 5/2 kernel with automatic relevance determination (ARD), which is known to improve convergence speed [23].

Due to the gradual spreading of the droplet over time, the experiments require time-efficient optimization strategies. To accelerate the Bayesian optimization process, several measures were implemented, including GPU Acceleration using a NVIDIA RTX 3060 GPU. Additionally, the optimization focuses solely on shaping the droplet, without adjusting its position, thereby simplifying the problem.

Given the continuous action space, there is an infinite number of possible actuation combinations for the eight actuators. Optimizing all actuators at once would be time-consuming, often leading to the droplet spreading over the workspace before convergence is reached. Therefore, batch Expected Improvement (qEI) is used as the acquisition function, balancing exploration and exploitation efficiently.

The optimization variables are bounded between 0 and 1, corresponding to 0 and 25 mT of magnetic flux density at the solenoid tip. The surrogate model is initially trained with approximately 10 informative actuation-response samples.

The acquisition function then selects the next batch of experiments. After each batch is executed, the new data is added to the dataset and the surrogate model is retrained.

The objective function is defined as the root mean square error (RMSE) between target shape ratios ($TSR_i$) and the response shape ratios ($RSR_i$), where $i = 1 \ldots 32$. The objective function for the optimization is given by:

$$J = \sqrt{\frac{1}{n}\sum_{i=0}^{n}(TSR_i - RSR_i)^2} \qquad (2)$$

### III. RESULTS

Before each experimental trial, the droplet is centered by actuating all eight solenoids with a 39 mT magnetic field at the tip for five seconds. The solenoids are then demagnetized using a three seconds exponentially decaying sinusoidal current signal given by $I = Ae^{-kt}\sin(\omega t)$, where amplitude $A = 2$ A, decay rate $k = 2.5$, and the angular frequency $\omega = 50\ rad\ s^{-1}$. This process ensures that the droplet adopts a circular shape prior to each trial. The solenoid actuation begins with a ramp-up signal to the desired magnetic field and is held constant for four seconds.

The shape response for primitive shapes with smoothed corners is shown in Figure 3. The droplet retains curvature at the corners due to surface tension, which limits the replication of sharp-cornered target shapes such as triangles and rectangles. Therefore, the corners of the target shapes are smoothed to improve compatibility. The Bayesian optimizer is initialized with five prior samples corresponding to the selected solenoids actuation patterns. Optimization is carried out in batches of five experiments, with each batch taking approximately 20 seconds to generate.

Figure 3a shows two optimization batches for shaping a triangle droplet with a height-to-base ratio (HBR) of 0.5. The sample number indicates the number of tests performed since initialization. Actuated solenoids are marked in green, and deactivated solenoids are marked in red.

The response for triangle shape commands with HBR of 0.50, 0.75, 1.0, 1.25, and 1.50 are shown in Figure 3c. Based on our previous experience with programmable droplet shaping at the air-ferrofluid interface [18], we selectively optimize magnetic fields for a subset of solenoids. Actuating a solenoid deforms the air-ferrofluid interface, pushing the droplet boundary away from the solenoid. Therefore, activating solenoids in directions where sharper corners are desired can be counterproductive. Deactivating these solenoids improves shape fidelity and accelerates convergence.

For the triangle with 0.5 HBR, only the NW, NE, and S solenoids are optimized. For higher HBR values, five solenoids, NW, NE, W, E, and S, are used. For rectangles with aspect ratios of 1.00, 1.25, 1.50, 1.75, and 2.00, the optimized solenoids are N, E, W, and S, as shown in Figure 3d. Figure 4a presents the optimization process for a triangle with HBR = 1.5 and a rectangle with an aspect ratio of 1.25. A supplementary video demonstrates the real-time optimization process for the triangle shape with HBR = 1.5.

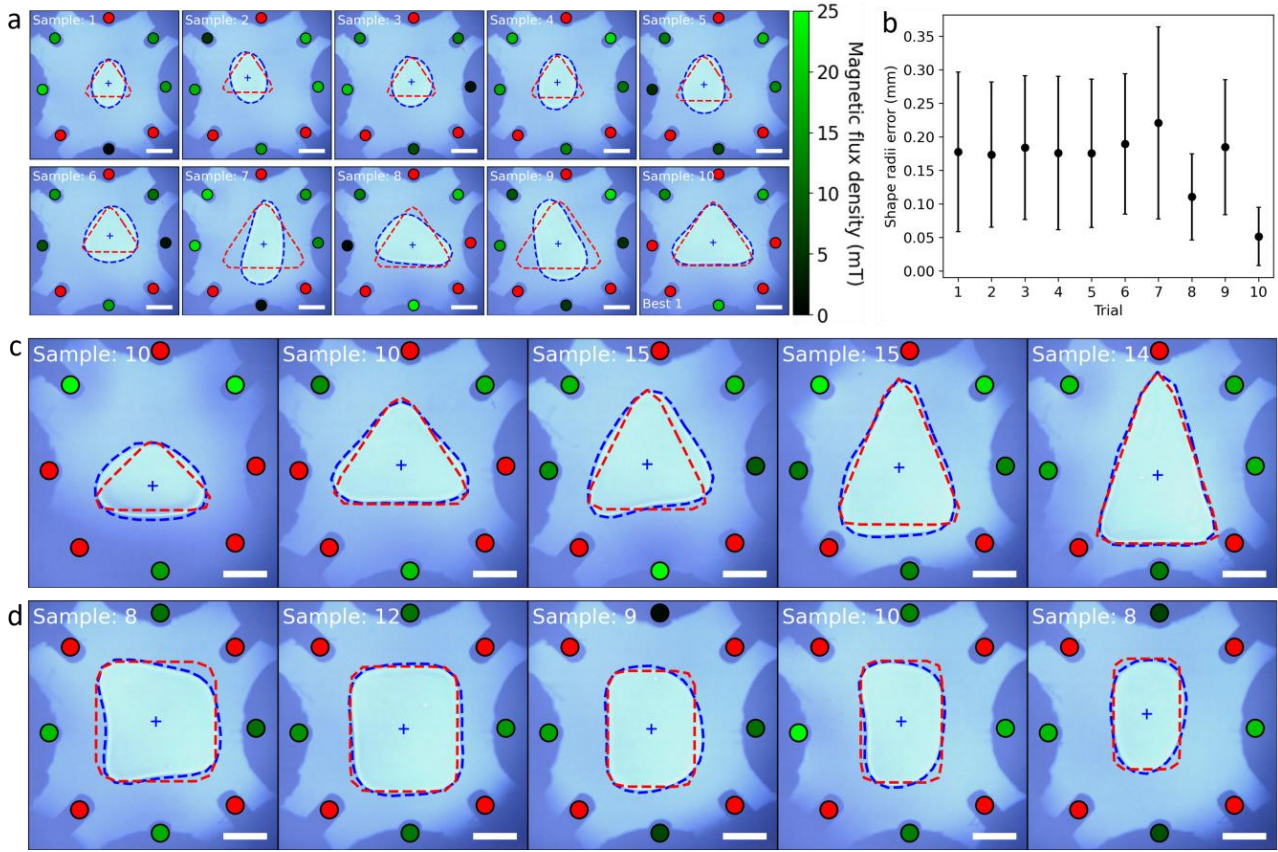

Figure 3. Shaping of a hexadecane droplet at the air-ferrofluid interface into primitive shape targets with curved corners. a) Data-efficient exploration process for a triangular shaped target. The green color bar represents the magnetic flux density of the actuated solenoids. The inactive solenoids are marked by red indicators. b) Shape radii error. c) Triangular shaping of droplets for different height to base ratios of the triangular target. d) Rectangular shaping of droplets for different aspect rations of the rectangle target. The scale bars are 2 mm.

The shape radii errors for triangle targets are 0.34 ± 0.17, 0.17 ± 0.14, 0.35 ± 0.22, 0.25 ± 0.17, 0.10 ± 0.08 mm, while for rectangular targets, the errors are 0.22 ± 0.16, 0.13 ± 0.08, 0.17 ± 0.12, 0.16 ± 0.12, 0.17 ± 0.11 mm. The maximum observed errors for triangle and rectangle shapes were 0.81 mm and 0.62 mm respectively. The shape errors are illustrated in Figure 4b-c.

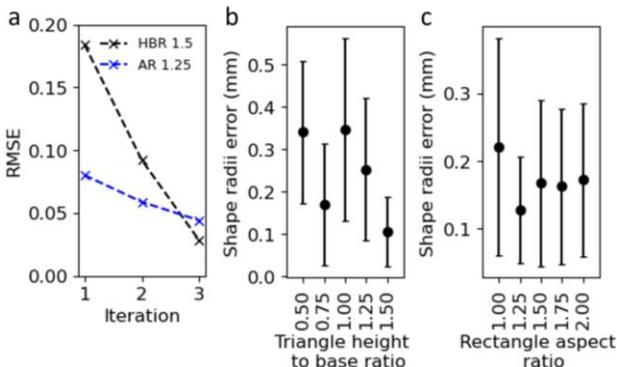

Figure 4. a) Optimization process for the triangle with a height to base ratio of 1.5 and the rectangle with an aspect ratio of 1.25. b) Shape radii error for triangular shapes and c) rectangular shapes.

These errors are attributed to several factors. First, surface tension and the interfacial physics of the air-liquid-liquid system inherently limit sharp-cornered shape formation. Secondly, the droplet gradually spreads and the interfacial chemical composition may change during shaping, making it difficult for the optimization algorithm to precisely predict required magnetic fields. Finally, the optimization process often cannot complete a sufficient number of iterations before the droplet spreads significantly over the workspace. While the shaping performance may not yet match the precision levels seen in highly controlled environments typically used in robotics and automation, the proposed method represents a significant advancement by enabling droplet manipulation without requiring magnetic droplet properties or predefined templates.

The optimization algorithm was also applied to shape the droplet into letters of "AALTO" as shown in Figure 5. Due to the geometric complexity of the letters, the algorithm would benefit from a large number of samples. However as described above, time constraints caused by droplet spreading limit the total number of experiments. To address this, only the most significant shape ratios, typically corresponding to corners, are selected and optimized to achieve the target shape. Flat areas are more easily maintained due to the surface tension.

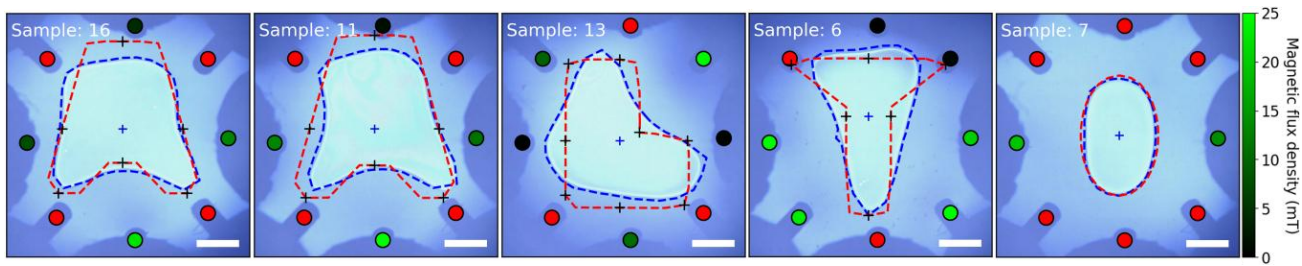

Figure 5. Shaping hexadecane droplets at the air-ferrofluid interface into letters of "AALTO". The red dashed lines denote the target shape. The blue dashed line is the shape response of the droplet. The black crosses denote the shape ratios to be optimized. The color bar represents the magnetic flux density of the actuated solenoids. Inactive solenoids are denoted by red indicators. The scale bars are 2 mm.

The optimized shape ratios are marked with a black cross in the figure. For the A shape, the N, E, W and S solenoids are optimized. For L, the NW, NE, W, E, and S solenoids are used. For T, all solenoids are used except S. For the O shape, which is relatively simple and elliptical, all 32 shape ratios were included in the objective function, and only the W and E solenoids are optimized.

The resulting droplet shape approximates the shape of the letter. Further improvement in the letter shaping may require shape-specific solenoid placement strategies.

## IV. CONCLUSION

In this study, hexadecane droplets at the air-ferrofluid interface were shaped into primitive shapes such as triangles and rectangles using a data-efficient exploration strategy based on Bayesian optimization method. The rapid convergence of Bayesian optimization helps minimize the influence of environmental factors such as temperature and humidity fluctuation. Using the proposed method, target shapes were achieved with a maximum radii error of 0.81 mm for primitive geometries.

The proposed method is both data-efficient and model-free, and can be applied to a wide range of droplet shaping systems regardless of the external energy field employed, e.g., air, acoustic, electric, or thermal. This flexibility enables integration into different setups without requiring extensive redesign.

In the current system, the spreading of the droplet at the air-ferrofluid interface limits the available experimental time. Without this constraint, the method could converge to target shapes more rapidly. Although each iteration of Bayesian optimization currently takes approximately 20 seconds, the use of faster or application specific GPUs in the future could significantly reduce this time and further enhance the convergence speed.

Future work may involve utilizing bio-compatible ferrofluids and extending this method to the manipulation of the shape of living organisms for tissue engineering and other biomedical applications.